\documentclass[11pt,dvips]{article}

\usepackage{epsfig,times}
\usepackage{picinpar}
\usepackage{wrapfig}
\usepackage{floatflt}
\makeatletter
\renewcommand{\section}{\@startsection{section}{1}{0in}
	{0.4\baselineskip}{0.1\baselineskip}{\Large\bf}}
\renewcommand{\subsection}{\@startsection{subsection}{2}{0in}
	{0.25\baselineskip}{-\baselineskip}{\large\bf}}
\renewcommand{\subsubsection}{\@startsection{subsubsection}{3}{0in}
	{0.1\baselineskip}{-\baselineskip}{\normalsize\bf}}
\makeatother
\begin{document}

%
%  Session and Paper Code:
\makeatletter\newcommand{\ps@icrc}{
\renewcommand{\@oddhead}{\slshape{HE.3.1.2}\hfil}}
\makeatother\thispagestyle{icrc}
%
%  ***INSTRUCTIONS:***  Replace `OG 9.9.9' in the command argument below
%			with your assigned session and paper code:
%\markright{HE 3.1.02}
%

%  Title:
\begin{center}
%
%  ***INSTRUCTIONS:***  Replace `Instructions for Preparation of Manuscript'
%			with your paper's title:
{\LARGE \bf Search for Steady, Modulated, and Variable Cosmic Ray Sources
Using Underground Muons in MACRO}
\end{center}

%  Author List:
\begin{center}
%
%  ***INSTRUCTIONS:***  Replace authors and addresses below with your own:
%
{\bf C. Satriano$^{1}$ for the MACRO Collaboration}\\
{\it $^{1}$Universit\'a della Basilicata, Potenza and INFN, Italy}
\end{center}

%  Abstract:
\begin{center}
{\large \bf Abstract\\}
\end{center}
\vspace{-0.5ex}
%
%  ***INSTRUCTIONS:***  Replace text below with your own abstract:
%
%
Using a sample of 38.5 million underground muons collected by the MACRO 
detector we have performed an all-sky search for pointlike sources
 producing excesses of muons above the expected background. The d.c. 
 muon flux upper limit at the Earth coming from selected sources is of the 
order of 10$^{-13}$ cm$^{-2}$ s$^{-1}$ or less. Futhermore we discuss searches 
for possible modulated and variable sources of muons using different
 techniques.

%  Leave this line skip in place:
\vspace{1ex}

%
%  Manuscript text:
%
%  ***INSTRUCTIONS:***  Delete the next few pages of text and enter your
%own.  There will
%			be a warning, `STOP DELETING TEXT!!', just before
%the References
%			section so that the standardized Reference heading
%will not be
%			accidently erased.  Within the text below is an
%example is given
%			of a figure placement (using `picinpar').
\section{Introduction:}
\label{intro.sec}
 In this work we present the results on the study of muon arrival 
directions, as seen by the MACRO detector (Ahlen et al., 1993), looking for 
excesses above the expected background in every sky direction. We have also 
searched for steady and pulsed signals from potential cosmic ray sources.

 Features of the MACRO detector are an excellent angular resolution and 
a large collecting area, when compared to other deep underground detectors. 
Because of these properties, MACRO can be used as a good muon telescope. 
The interest in this kind of analysis arises from the old observation of 
underground muon excesses from the binary system Cygnus X-3, as reported from 
the SOUDAN and NUSEX collaborations (Marshak et al., 1985; Battistoni et al.,
 1985).
On 1991 January, SOUDAN2 reported the observation of unmodulated muon signals 
from Cyg X-3, correlated with the radio-burst occurring in the same period. 
These muon excesses had fluxes of 7.5$\times$$10^{-10}$$cm^{-2}s^{-1}$ 
(Marshak, 1993). Recently, no experiment detected signals from this kind of 
source, but several observations of X and $\gamma$-bursts were reported from 
other celestial objects. During 1997, several $\gamma$-bursts from AGN objects,
like Markarian 421 and Markarian 501, were reported by the Whipple and Hegra
experiments (Punch et al., 1992; Quinn, 1997). 

\section{Data Selection and Reduction:}
\label{data.sec}
 The data set used for this analysis is from a long period of acquisition, 
from the first MACRO run (February 1989) until January 1999. 
In order to optimize the
track reconstruction and to eliminate possible detector malfunctions,
strict selection criteria were defined. 
After all cuts, we selected a
sample of 38.5 million  events (single and doubles), collected over 55248 h 
of live time.

\subsection{Angular Resolution and Background Simulation:}
\label{angular.sec}
 The MACRO pointing precision was calculated from the distribution of the
double-$\mu$ angular deviation. We assume that these events at the Earth's 
surface are parallel muon pairs. Then, the overburden rock produces their
deviation at the MACRO  level through multiple Coulomb scattering processes.
The space angle $\theta$, containing 68\% of the events, defines the
detector angular resolution. We found $\theta$ = $0.8^{o}$, consistent with the
value derived from the Moon shadowing of primary cosmic rays (Giglietto, 1999).

 For the present search, we assume that our signals are muons produced in
 atmospheric 
photoproduction processes and the background (or \emph{noise}) is
produced by primary interactions with the terrestrial atmosphere. 
The background
has been determinated by a Monte Carlo simulation, assuming an isotropically
distributed cosmic ray flux. For each run we generated the
background events by coupling the arrival direction of each muon with 25
arrival times, randomly extracted from the same run. 

\section{All-sky d.c. Survey:}
\label{all-sky.sec}

In the search for point-like sources we first started an all-sky survey, 
without
$\emph{a priori}$ assumptions about the source locations; then, we examined
selected celestial objects known as potential cosmic ray sources. 
Therefore, we examine sky regions looking for deviations from the expected 
background larger than random fluctuations.
 
\subsection{Search for d.c. Muon Excesses:}
As a first step, we generated the sky map with muons collected by MACRO by
dividing  the sky into bins of equal solid angle ($\Delta$$\Omega$= 2.1
$\times$ $10^{-3}$ sr; $\Delta$$\alpha$ = $3^o$, $\Delta$sin$\delta$ = 0.04). 
\begin{wraptable}{r}{10cm}
\begin{center} %da togliere se non va bene

\begin{tabular}{|c|c|c|c|c|c|c|}
\hline
		& MEAN               &  SIGMA  & $\chi^{2}$/Dof &N.deviations \\
\hline\hline
\emph{map(1)}   &-0.13$\times10^{-1}$&  0.99   &   62/35        &4(+)   3(-)  \\
\hline
\emph{map(2)}   &-0.6$\times10^{-3}$ &  1.01   &   24/34        &5(+)   3(-)  \\
\hline
\emph{map(3)}   &-0.22$\times10^{-3}$&  1.01   &   36/34        &4(+)   4(-)  \\
\hline          		
\emph{map(4)}   & 0.3$\times10^{-2}$ &  1.02   &   34/35        &5(+)   4(-)  \\
\hline
\end{tabular}
\end{center} %da togliere se non va bene
\caption{\emph{Best-fit parameters of the Gaussian fit for the four sky 
surveys.The last columns gives the number of positive and negative deviations 
larger than 3.2$\sigma$. }} 
%%\vspace*{0.8truecm}
\end{wraptable}
These bins have the same $\Delta$$\Omega$ as a narrow cone of
half-angle $1.5^{o}$.
Then, assuming that a potential source is located at the center of the bin, we
obtained a sky survey in absolute astronomical coordinates $\alpha$ and
sin$\delta$ ($\emph{map(1)}$ in Table 1).
However, if the source is near a bin edge, its signal will be spread between
adjacent bins. Then, to reduce the possibility of missing sources close to the
bin edges, three
other surveys were also done: $\emph{map(2)}$ shifted the grid one-half bin
in $\alpha$, $\emph{map(3)}$ shifted the grid one-half bin in sin$\delta$ and
$\emph{map(4)}$ which contains the $\emph{map(2)}$ and $\emph{map(3)}$ shift.
 For each solid angle bin in the four surveys, we calculated the deviation
from the mean in units of standard deviations: 
\begin{equation}
\mathop{\mathrm{\sigma}}(i)=
\frac{\displaystyle N_{obs}(i)-N_{exp}(i)}
{\displaystyle \sqrt{N_{exp}(i)}\qquad}\\ \\ 
\end{equation}
\\  
where $N_{obs}(i)$ is the observed number of events in the bin and 
$N_{exp}(i)$
is the number of events expected from the background simulation.
Bins having less than 20 events were removed.

 If the observed muons are distributed according to Gaussian statistics, the
$\sigma(i)$ distributions will be a Gaussian curve with zero mean and unit
 standard deviation. 
Table 1 shows the Gaussian best fit for the four surveys: the means are close to 
zero and the half widths near one. The best-fit parameters and the number of
fluctuations
larger than 3.2$\sigma$ are also given. We found the larger positive and 
negative deviations are in similar number. 
We conclude that all deviations are consistent with random background
fluctuations. 

\subsection{Steady Flux Limits:}
\label{steady.sec}
We calculated the upper limits to the muon flux from the all-sky survey, 
following the
Helene method (Helene, 1983). The steady upper limit at 95\% confidence 
level to the muon flux for each bin was computed from: 
\begin{equation}
\mathop{\mathrm J^{stdy}_{\mu}(95\%)}\leq
\frac{\displaystyle n_{\mu}(95\%)}{\displaystyle KA_{eff}T_{exp}}
\mathop{\mathrm{cm^{-2}s^{-1}}}.\\ \\
\end{equation}
Here $n_{\mu}$(95\%) represents the muon number we would observe at 95\%
C.L. in each bin, if $N_{obs}(i)$ and $N_{exp}(i)$ are  the
number of events observed and expected in that bin, respectively. 
$A_{eff}$ is the average effective area for every bin computed by
averaging the projected area seen by each muon
\begin{equation}
\mathop{\mathrm A_{eff}}(i)=
\frac{\displaystyle 1}{N_{obs}(i)}
\sum_{j=1}^{N_{obs}(i)}A({\alpha}_{i},{\delta}_{i})
\end{equation}
where ${\alpha}_{i}$ and ${\delta}_{i}$ are the muon arrival directions. The
$A_{eff}$ calculation takes into account the geometrical and tracking
reconstruction efficiencies.
$T_{exp}$ is the exposure time, computed bin-by-bin 
and the factor K takes into account the scattering of some muons out of the
bin. 
We found K=0.78 for the selected bin dimension ($1.5^{o}$ half-angle).

We found, for the majority of the bins, 
$J_{\mu}^{stdy}$(95\%) $\leq$ 5$\times$$10^{-13}cm^{-2}s^{-1}$.

\subsection{Search for Modulated Signals:}
\label{modulated.sec}
Searches were also made for modulated signals coming from those sources that 
in the past showed variability (Cyg X-3 and Her X-1). We found no evidence 
for excesses in any of the phase bins into which the characteristic period was
 divided. The 95\% C.L. upper limits to the modulated muon flux from the 
directions of Cyg X-3 and Her X-1 are, respectively,  
1.8$\times$$10^{-13}cm^{-2}s^{-1}$ and 2.2$\times$$10^{-13}cm^{-2}s^{-1}$.

\section{Point-like d.c. Source Study:}
\label{point.sec}
\begin{wraptable}{r}{12.5cm}
\begin{center}
\begin{tabular}{|c|c|c|c|c|c|c|c|}
\hline
       &         &         &        &Area  &$T_{exp}$&    Flux 1     &   Flux 2   \\
       &$N_{obs}$&$N_{exp}$&$\sigma$&      & $10^6$  &               &              \\
       &         &         &        &$m^2$ &  sec    &$cm^{-2}s^{-1}$&$cm^{-2}s^{-1}$\\
\hline\hline                         
Cyg X-3&  4242   & 4234.2  &  -0.1  & 691  &   109   & 3.4 $10^{-13}$& 1.9 $10^{-13}$\\
\hline
Mrk 421&  4328   & 4250.4  &   1.2  & 704  &   105.4 & 4.4 $10^{-13}$& 2.0 $10^{-13}$\\
\hline 
Mrk 501&  4630   & 4496.1  &   2.0  & 699  &   107.4 & 5.4 $10^{-13}$& 2.0 $10^{-13}$\\
\hline          		
\end{tabular}
\end{center}
\caption{\emph{Number of events observed and expected, the standard deviation
calculated as in (1), the average effective 
area and exposure time for three selected sources. For these we give also 
the flux upper limits calculated with the Helene formula (Flux 1) and with the 
classical method (Flux 2).}} 
\end{wraptable}

To study muons coming from specific celestial objects, we selected
only events contained in a narrow cone ($1.0^o$ half-angle) around the source 
direction. The selected windows are centered on the Cyg X-3, Mrk 421 and 
Mrk 501 positions. Table 2 gives the muon number observed and expected from the
 Monte Carlo simulation. 
Since there is no significant excess from Cygnus X-3, we calculated its steady 
flux limit following the Helene formula (Helene, 1983) and the classical 
method (Hikasa et al., Particle Data Group, 1992).
 
For Mrk 421 and Mrk 501 we have small steady excesses at the level 
of 1.2$\sigma$ and 2.0$\sigma$ respectively. For these sources the steady flux 
limits are given in Table 2.

\section{Search for Bursting Episodes:}
\label{bursts.sec}

\begin{wrapfigure}{r}{7cm}
    \epsfxsize=7cm
    \centerline{\epsffile{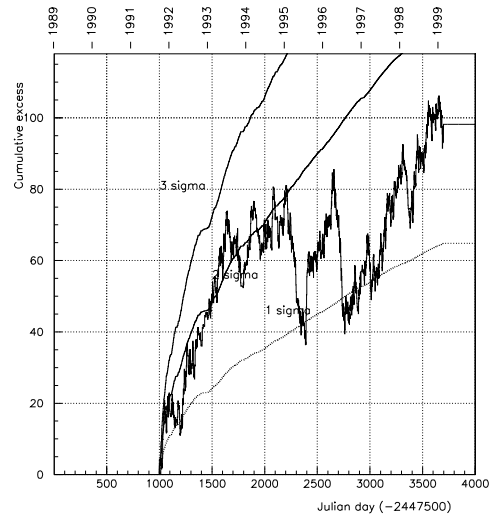}}
  \caption{Cumulative muon excesses from the direction of Mrk 501 
($1^{o}$ half-angle). The selected time period is from January 1992 until 
January 1999}
\end{wrapfigure}

During 1997, surface experiments reported several $\gamma$-bursts from
celestial objects like Mrk 421 and Mrk 501.
Therefore we have made also a search for pulsed muon signals in a narrow window 
($1^o$ half-angle), around the position of these possible 
sources of U.H.E. photons. 
We studied bursting episodes following two different methods.

 First, we searched for daily excesses of muon flux above the
background: because the daily observed events are few, we plotted the
cumulative excesses day by day, over a long observation period (Clay, 
Dawson, \& Meyhandan, 1994).
Fig.1 shows the progressive accumulation of our
excesses above the computed background for Mrk 501.
Periods are present in which excesses accumulated rapidly but after that, the
cumulation rate reduced.

 We observe the cumulated signal is every day above the
background: fluctuations are contained in 2$\sigma$ for Mrk 421 and for Mrk 501.

 For the same sources, we have also searched for statistically significant
daily excesses of muons, following a second method (Padilla et al., 1998).
From our complete data set 
we computed, day by day, the 
quantity $-log_{10}$ P, where:
\begin{equation}
\mathop{\mathrm {P}}= 1- 
\sum_{n=0}^{N_{obs}-1}   
\frac {\displaystyle \alpha^n}{\displaystyle (1+\alpha)^{n+N_{bck}+1}\qquad}
\frac {\displaystyle (n+N_{bck})!}{\displaystyle n!\times N_{bck}!\qquad}
\end{equation}
$N_{obs}$ is the number of observed muons in a day and $N_{bck}$ is the number
 of background events in the same day, computed by Monte Carlo simulation. 
 $\alpha$ is the ratio of the on-source time to the off-source time (0.04 for 
the present analysis) and P 
represents the probability of observing a burst at least as large as $N_{obs}$
due to background fluctuations. We assume that the background has a Poissonian
distribution about $N_{bck}$. If no bursts are observed, we expect that the
cumulative frequency distribution of P is a power law of index -1.
In Table 3 we report the statistical parameters for Mrk421 to be compared 
with the observations from  surface experiments 
(Shubnell et al., 1996). For comparison, we created  a sky survey by 
computing the quantity -$log_{10}$ P for every bin of the sky map:
this distribution has slope -1.1 and the superposed line fits the data
well. 

\section{Conclusions:}
\label{conclusions.sec}
\begin{wraptable}{r}{8.2cm}
\begin{center}
\begin{tabular}{|c|c|c|c|c|}
\hline
%         &         &         &                   &                     \\
 Date    &$N_{obs}$&$N_{exp}$&$-log_{10}P$ &           P         \\
%         &         &         &                   &                     \\
\hline\hline                         
 7 Jan 93 &   9     &  1.88   &      3.58         &2.6$\times$$10^{-4}$    \\
\hline
14 Feb 95 &   9     &  1.96   &      3.46         &3.6$\times$$10^{-4}$    \\
\hline 
27 Aug 97 &   8     &  1.72   &      3.16         &6.9$\times$$10^{-4}$    \\
\hline          		
5 Dec 98  &   8     &  1.52   &      3.48         &3.3$\times$$10^{-4}$     \\
\hline          		
\end{tabular}
\end{center}
\caption{\emph{Statistical parameters for possible excesses 
from Mrk421. $N_{exp}=\alpha \times N_{bck}$ and P represents the probability
per day to observe a burst at least as 
large as $N_{obs}$, due to background fluctuations ($1^{o}$ half-angle). The 
total number of observations was 3660.}} 
\end{wraptable}

 Since February 1989, the MACRO detector has collected a sample of about
38.5 million muons. Using this sample, we searched for muon excesses above
background from every sky direction and from single cosmic ray sources.
No significant excesses have been found from the all-sky survey and  we 
computed the upper steady limit to the muon flux. At 95\% confidence level, we 
found this limit at  5$\times$$10^{-13}$ $cm^{-2}s^{-1}$.
The search for steady emission from different sources does not indicate for
Mrk 421 and Mrk 501 a  muon excess above the background, as the 
%statistical significance, 2.1$\sigma$ and 3$\sigma$ respectively.
statistical significance of the excess is 1.2$\sigma$ and 2$\sigma$ 
respectively.

%\vspace*{4.5truecm}

%\newpage
%
%  References: (DO NOT ALTER NEXT 4 LINES)
\vspace{1ex}
\begin{center}
{\Large\bf References}
\end{center}
%
%  ***INSTRUCTIONS:***  Enter your references alphabetically following the
%format of the example citations below.
Ahlen, S. et al. 1993, ApJ 412, 301\\
Battistoni, G. et al. 1985, Phys. Rev. Lett. B 160, 465\\  
Clay, R.W., Dawson, B.R., Meyhandan, R. 1994, Astrop. Phys. 2, 347\\
Giglietto, N.,(MACRO Collaboration), SH.3.2.38,Proc. 26th ICRC (Salt Lake City, 1999)\\
Marshak, M.F. et al. 1985, Phys. Rev. Lett. B 54, 2079\\
Marshak, M.L. Proc. $23^{rd}$ ICRC (Calgary, 1993), 4, 458\\
Helene, O. 1983, Nucl. Instr. Meth. 212, 319\\
Hikasa, K. et al. (Particle Data Group), 1992, Phys. Rev. D45, S1\\
Padilla, L. et al. 1998 A\&A 337,43\\ 
Punch, M. et al., 1992, Nature 358, 477\\
Quinn, J. Proc. $25^{th}$ ICRC (Durban, 1997), 3, 249\\
Shubnell, M.S. et al. 1996, ApJ 460, 644\\
\end{document}